\documentclass[aps,prl,twocolumn,showpacs]{revtex4}
\usepackage{times,bbm,amsmath,amssymb,amsbsy,graphicx,subfigure}

\usepackage[latin1]{inputenc}

\newcommand{\gr}[1]{\boldsymbol{#1}}
\newcommand{\sig}{\gr\sigma}
\newcommand{\be}{\begin{equation}}
\newcommand{\ee}{\end{equation}}
\newcommand{\bea}{\begin{eqnarray}}
\newcommand{\eea}{\end{eqnarray}}

\newcommand{\N}{{\cal N}}

\newcommand{\R}{\mathbbm R}

\renewcommand{\det}{{\rm Det}\,}
\newcommand{\eq}[1]{Eq.~(\ref{#1})}

\begin{document}
\title{Quantification and scaling of multipartite entanglement in continuous variable systems}
\date{September 1, 2004}
\author{Gerardo Adesso}
\author{Alessio Serafini}
\author{Fabrizio Illuminati}
\affiliation{Dipartimento di Fisica ``E. R. Caianiello'',
Universit\`a di Salerno, INFM UdR di Salerno, INFN Sezione di Napoli,
Gruppo Collegato di Salerno,
Via S. Allende, 84081 Baronissi (SA), Italy}
\begin{abstract}
We present a theoretical method to determine the multipartite
entanglement between different partitions of multimode,
fully or partially symmetric Gaussian states of continuous
variable systems. For such states, we determine the exact 
expression of the logarithmic negativity and show that it
coincides with that of equivalent two--mode Gaussian states.
Exploiting this reduction, we demonstrate the scaling of the 
multipartite entanglement with the number of modes and its 
reliable experimental estimate by direct measurements of the 
global and local purities.
\end{abstract}

\maketitle

The full understanding of the structure of multipartite quantum entanglement
is a major scope in quantum information theory that is yet to be achieved.
At the experimental level, it would be crucial to devise
effective strategies to conveniently distribute the entanglement between different
parties, depending on the needs of the addressed information protocol.
Concerning the theory, the conditions of separability for generic bipartitions
of Gaussian states of continuous variable (CV) systems have been derived and
analysed \cite{ngau1,ngau2,vloock03}.
However, the quantification and scaling of entanglement for arbitrary
states of multipartite systems remains in general a formidable task \cite{vloock02}.
In this work, we present a theoretical scheme to exactly determine the multipartite
entanglement of generic Gaussian symmetric states (pure or mixed) of CV systems.

We consider a CV system consisting of $N$ canonical bosonic modes,
associated to an infinite-dimensional Hilbert space and described by the
vector $\hat{X}$ of the field quadrature operators. Quantum states
of paramount importance in CV systems are the so-called Gaussian
states, {\em i.e.}~states fully characterized by first and second
moments of the canonical operators. When addressing physical
properties invariant under local unitary transformations, one can
neglect first moments and completely characterize Gaussian states by
the $2N \times 2N$ real covariance matrix (CM) $\gr{\sigma}$, whose
entries are $\sigma_{ij}=1/2\langle\{\hat{X}_i,\hat{X}_j\}\rangle
-\langle\hat{X}_i\rangle\langle\hat{X}_j\rangle$. The CM
$\gr{\sigma}$ must fulfill the uncertainty relation
$\gr{\sigma}+i\Omega \geq 0$, with the symplectic form
$\Omega=\oplus_{i=1}^{n}\omega$ and $\omega=\delta_{ij-1}-
\delta_{ij+1},\, i,j=1,2.$ Symplectic operations ({\em
i.e.}~belonging to the group $Sp_{(2N,\R)}= \{S\in
SL(2N,\R)\,:\,S^T\Omega S=\Omega\}$) acting by congruence on CMs in
phase space, amount to unitary operations on density matrices in
Hilbert space. In phase space, any $N$-mode Gaussian state can be
written as $\gr{\sigma}= S^T \gr{\nu} S$, with $\gr{\nu}=\,{\rm
diag}\,\{n_1,n_1,\ldots n_N, n_N \}$. The set
$\Sigma=\{n_i\}$ constitutes the symplectic spectrum of
$\gr{\sigma}$ and its elements must fulfill the conditions $n_i\ge
1$, ensuring positivity of the density matrix $\varrho$ associated to
$\gr{\sigma}$. The symplectic eigenvalues $n_i$ can be computed as
the eigenvalues of the matrix $|i\Omega\gr{\sigma}|$.
They are determined by $N$ symplectic invariants associated to the
characteristic polynomial of such a matrix: two global invariants which
will be useful are the determinant $\,{\rm
Det}\,\gr{\sigma}=\prod_{i}n_i^2$ and the seralian $\Delta=\sum_i
n_i^2$, which is the sum of the determinants of all the $2\times
2$ submatrices of $\gr{\sigma}$ related to each mode.

The degree of mixedness of a quantum state $\varrho$ is
characterized by its purity $\mu=\,{\rm Tr}\,\varrho^2$. For a
Gaussian state with CM $\gr{\sigma}$ one has simply
$\mu=1/\sqrt{\,{\rm Det}\,\gr{\sigma}}$. As for the entanglement, we
recall that positivity of the partially transposed state $\tilde{\varrho}$,
obtained by transposing the reduced state of only one of the subsystems,
is a necessary and sufficient condition (PPT criterion) 
of separability for $(N+1)$-mode Gaussian states of $1\times N$-mode partitions 
\cite{simon00,werner02}. In
phase space, partial transposition amounts to a mirror reflection of
one quadrature associated to the single-mode partition. If
$\{\tilde{n}_i\}$ is the symplectic spectrum of the partially
transposed CM $\tilde{\gr{\sigma}}$, then a $(N+1)$-mode Gaussian
state with CM $\gr{\sigma}$ is separable if and only if
$\tilde{n}_i\ge 1$ $\forall\, i$. A proper measure of CV
entanglement is the logarithmic negativity $E_{\N}$ \cite{vidal02},
which is readily computed in terms of the symplectic spectrum
$\tilde{n}_i$ of $\tilde{\gr{\sigma}}$ as
$E_{\N}=-\sum_{i:\tilde{n}_i<1}\ln \tilde{n}_i$. Such a measure
quantifies the extent to which the PPT condition $\tilde{n}_i\ge 1$
is violated.

Let us first consider the $2N \times 2N$ CM $\sig_{\beta^N}$ of a fully symmetric $N$-mode
Gaussian state ({\em i.e.}~ a state invariant under the exchange of any two modes)

\vspace*{-.5cm}
\begin{equation}\label{fscm}\hspace*{1cm}
\sig_{\beta^N}={\left(%
 \begin{array}{cccc}
  \gr\beta & \gr\varepsilon & \cdots & \gr\varepsilon \\
  \gr\varepsilon & \gr\beta & \gr\varepsilon & \vdots \\
  \vdots & \gr\varepsilon & \ddots & \gr\varepsilon \\
  \gr\varepsilon & \cdots & \gr\varepsilon & \gr\beta \\
\end{array}%
\right)}\,,
\end{equation}
where $\gr\beta$ and $\gr\varepsilon$ are $2\times2$
submatrices.
Due to the symmetry of such a state, $\gr\beta$ and $\gr\varepsilon$
can be put by means of local (single-mode) symplectic operations in the form
$\gr\beta={\rm diag}\{b,\,b\}, \; \gr\varepsilon={\rm
diag}\{e_1,\,e_2\}$.
The symplectic spectrum $\Sigma_{\beta^N}$ of $\sig_{\beta^N}$
has then the structure (see the Appendix)
\begin{eqnarray}\label{fssymp}
\Sigma_{\beta^N}&=&\{\underbrace{\nu_-,\,\ldots,\nu_-}_{N-1},\,\nu_{+^{(N)}}\}\,,\\
\nu_-^2 &=& (b-e_1)(b-e_2)\,, \nonumber \\
\nu_{+^{(N)}}^2 &=& (b+(N-1)e_1)(b+(N-1)e_2)\,. \nonumber
\end{eqnarray}
The $(N-1)$-degenerate eigenvalue $\nu_-$ is independent of $N$,
while $\nu_{+^{(N)}}$ can be expressed as a function of the
purity $\mu_\beta\equiv(\det\gr\beta)^{-1/2}$ of the single--mode 
reduced state and of the
symplectic spectrum of the two-mode block $\sig_{\beta^2}$,
$\Sigma_{\beta^2}=\{\nu_-,\,\nu_+ \equiv \nu_{+^{(2)}}\}$
\begin{equation}\label{fsnupiun}
\nu_{+^{(N)}}^2=-\frac{N(N-2)}{\mu_\beta^2}+\frac{(N-1)}2\big( N
\nu_+^2+(N-2)\nu_-^2 \big)\,.
\end{equation}
The global purity of the fully symmetric state is
\begin{equation}\label{fsmu}
\mu_{\beta^N}\equiv\left(\det\sig_{\beta^N}\right)^{-1/2}=
\left(\nu_-^{N-1} \nu_{+^{(N)}}\right)^{-1}\,,
\end{equation}
and, through \eq{fsnupiun}, can be directly linked to the one- and
two-mode parameters. In particular,
the symplectic eigenvalues $\nu_\mp$ are determined in terms
of the two $Sp_{(4,\mathbbm{R})}$ invariants $\mu_{\beta^2}$ and
$\Delta_{\beta^2}$ by the relation \cite{adesso04}: 
$2\nu_\mp^2 = \Delta_{\beta^2}\mp\sqrt{\Delta_{\beta^2}^2-4/\mu_{\beta^2}^2}$.

Next, we consider the $(N+1)$-mode Gaussian states 
constituted by generic single-mode states with CM $\gr\alpha$
and fully symmetric $N$-mode states with CM $\sig_{\beta^N}$ of the form 
(\ref{fscm}). The mode with CM $\gr\alpha$ is then coupled with all other $N$
modes by the same $2 \times 2$ real matrix $\gr\gamma$. The CM $\sig$ of such
$(N+1)$-mode states reads
\begin{equation}\label{sig}
\hspace*{-.1cm}
\sig\!=\!\left(%
\!\begin{array}{cc}
  \gr\alpha & \gr\Gamma \\
  \gr\Gamma^{\sf T} & \sig_{\beta^N} \\
\end{array}%
\!\!\right)\!,\;\;\gr\Gamma\equiv(\underbrace{\gr\gamma\;\cdots\;\gr\gamma}_N) \; .
\end{equation}
We will now show that the properties of mixedness and entanglement of these states 
are determined by a suitable, limited set of global and local invariants under 
symplectic (unitary) operations.
Let us introduce the purity $\mu_\alpha \equiv (\det\gr\alpha)^{-1/2}$ 
of the single-mode party, the global purity $\mu_{\sigma} \equiv (\det\sig)^{-1/2}$ 
of the state (\ref{sig}), and the global
$Sp_{(2N+2,\mathbbm{R})}$ invariant $\Delta_\sigma \equiv \Delta_{\alpha\gamma} + 
\Delta_{\beta^N} = \sum_{i} n_i^2$, where the
$n_i$'s constitute the symplectic spectrum
$\Sigma=\{n_1,\,\ldots,n_{N+1}\}$ of the CM $\sig$, and
\begin{eqnarray}
  \Delta_{\alpha\gamma} &\equiv& \det\gr\alpha+2N\det\gr\gamma\,, \label{deltaa} \\
  \Delta_{\beta^N} &\equiv& N (\det\gr\beta+(N-1)\det\gr\varepsilon) \; .
  \label{deltab}
\end{eqnarray}
We are now in the position to characterize and quantify the bipartite entanglement
between the single mode $\gr\alpha$ and the $N$-mode block 
$\sig_{\beta^N}$, the multipartite entanglement between all the
$N+1$ modes, and to provide an operational scheme for their 
experimental determination
in terms of measurements of the global and local purities.
To proceed, we must evaluate
the logarithmic negativity by
determining the partially transposed CM $\tilde{\gr{\sigma}}$,
with respect to the partition $\alpha\vert\beta^N$, which is obtained 
by flipping the sign of $\det\gr\gamma$. Mixedness and
entanglement are encoded respectively in the symplectic spectrum of
$\sig$, and of $\tilde{\sig}$. It
is worth noting that, of the previously introduced parameters,
only $\Delta_{\alpha\gamma}$ is affected by the operation of partial
transposition: $
\Delta_{\alpha\gamma}
\overset{\sig\rightarrow\tilde{\sig}}{\longrightarrow}\tilde{\Delta}_{\alpha\gamma}$,
with
\be
\tilde{\Delta}_{\alpha\gamma} \equiv \det\gr\alpha-2N\det\gr\gamma \equiv
-\Delta_{\alpha\gamma} + 2/\mu_\alpha^2 \; . \label{deltaapt}
\ee
The symplectic spectrum $\Sigma=\{n_i\}$
($i=1,\ldots,N+1$) of the CM 
$\sig$ \eq{sig} is
of the form (see Appendix)
\begin{equation}\label{sympn}
\Sigma=\{\underbrace{\nu_-,\,\ldots,\,\nu_-}_{N-1},\,
n_{-},\,n_{+}\}\,,
\end{equation}
where $\nu_-$ is the lowest symplectic eigenvalue of the
reduced two-mode state $\sig_{\beta^2}$. The eigenvalues $n_{\mp}$ can
be evaluated observing that
Eqs.~(\ref{fsmu},\ref{deltab},\ref{sympn})
impose the identity
$\Delta_\sigma\!\!=\!\!
\Delta_{\alpha\gamma} +(N-1)\nu_-^2+(\nu_-^{N-1}\mu_{\beta^N})^{-2}$ 
which can be used to obtain
\begin{eqnarray}\label{bisymp}
\hspace*{-2cm}
2n_{\mp}^2&=& \Delta_{\alpha\gamma} + (\nu_-^{N-1}\mu_{\beta^N})^{-2}
\nonumber \\
&\mp&
\sqrt{\big(\Delta_{\alpha\gamma} + (\nu_-^{N-1}\mu_{\beta^N})^{-2}\big)^2
-\frac4{(\nu_-^{N-1}\mu_{\sigma})^2}} \; .
\end{eqnarray}
Since partial
transposition leaves the $N$-mode symmetric block $\sig_{\beta^N}$
unchanged, the symplectic eigenvalues of $\tilde\sig$ are again of
the form
$\tilde\Sigma\equiv\{\tilde{n}_i\}=\{\nu_-,\,\ldots,\,\nu_-,\,
\tilde{n}_{-},\,\tilde{n}_{+}\}$, with $\tilde{n}_{\mp}$ defined as
in \eq{bisymp}, but with $\Delta_{\alpha\gamma}$ replaced by
$\tilde{\Delta}_{\alpha\gamma}$ from \eq{deltaapt}. The logarithmic negativity
$E_\N^{\alpha\vert\beta^N}$, quantifying the bipartite entanglement
between $\gr\alpha$ and $\sig_{\beta^N}$, is determined only by
those symplectic eigenvalues of $\tilde\sig$ which satisfy
$\tilde{n}_{i} < 1$. Since $\nu_- \ge 1$ (because it  belongs to the
symplectic spectrum of $\sig$), the entanglement is determined only
by the eigenvalues $\tilde{n}_{\mp}$. On the other hand, the
eigenvalues $n_{\mp}$ of \eq{bisymp} can be interpreted as the
symplectic spectrum of an {\em equivalent} two--mode state of CM
$\sig^{eq}$ with global purity $\mu^{eq}$ and seralian $\Delta^{eq}$ 
given by
\begin{equation}
  \mu^{eq} \equiv \nu_-^{N-1}\mu_\sigma\,, \qquad \Delta^{eq}
  \equiv \Delta_{\alpha\gamma} + (\nu_-^{N-1}\mu_{\beta^N})^{-2}\,.
  \label{mueq}
  \end{equation}
The corresponding $\tilde{\Delta}^{eq}$ associated to the partially
transposed CM $\tilde{\sig}^{eq}$ reads then $
\tilde{\Delta}^{eq}\equiv
-\Delta^{eq}+2/{\mu_\alpha^2}+2/{(\nu_-^{N-1}\mu_{\beta^N})^2} $. By
comparison with the expression
$\tilde{\Delta}=-\Delta+2/\mu_1^2+2/\mu_2^2$, holding for a generic
two--mode state with local purities $\mu_1$ and $\mu_2$ \cite{adesso04}, 
one determines the local purities of the equivalent two--mode 
state $\sig^{eq}$:
\begin{equation}\label{muieq}
  \mu_1^{eq} = \mu_\alpha\,,\qquad
  \mu_2^{eq} = \nu_-^{N-1} \mu_{\beta^N}\,.
\end{equation}
The two global invariants Eq.~(\ref{mueq}) and the two local
invariants Eq.~(\ref{muieq}) determine uniquely the entanglement
of the two--mode Gaussian state with CM $\sig^{eq}$.
In particular, it is immediate to see that the symplectic eigenvalues
of the partially transposed CM $\tilde{\sig}^{eq}$ coincide with  
$\tilde{n}_{\mp}$, so that we obtain the crucial result that
the logarithmic negativity of the equivalent two--mode state coincides
with the logarithmic negativity $E_\N^{\alpha\vert\beta^N}$ of the
$(N + 1)$--mode state. Explicitely, one has:
\begin{equation}
E_\N^{\alpha\vert\beta^N}=\max\{0,\,-\log\tilde{n}_-\} \; ,
\label{crucialone}
\end{equation}
with $\quad 2\,\tilde{n}_-^2 \equiv \
\tilde\Delta^{eq} - \sqrt{{\tilde\Delta^{^2eq}}-4/{\mu^{^2eq}}}$.
Indeed, only the smallest symplectic eigenvalue $\tilde{n}_-$ enters
in the determination of the multimode entanglement, since 
$\tilde{n}_{+} > 1$ for two-mode states \cite{adesso04}.

The $1 \times N$ entanglement is completely
quantified by measuring the two local purities $\mu_\alpha$ and $\mu_{\beta^N}$,
the global purity $\mu_\sigma$, the symplectic eigenvalue $\nu_{-}$, and $\det \gr\gamma$ (which
together with $\mu_{\alpha}$ determines $\Delta_{\alpha\gamma}$). The experimental determination
of these five quantities requires the full homodyne reconstruction of the $(N+1)$--mode CM \eq{sig}.
On the other hand,
the study of the entanglement of two-mode Gaussian states has shown
that a reliable quantitative estimate of the logarithmic negativity,
yielding exact (and very narrow) lower and upper bounds on the
entanglement, can be obtained by simply measuring the global and
local purities of the state \cite{adesso04}. In the present
instance, this fact implies that a reliable estimate of the $1 \times N$ 
entanglement does not require the knowledge of the correlation matrix 
$\gr\gamma$, while the remaining four quantities (the three purities and
the eigenvalue $\nu_{-}$) can be measured even without homodyning by direct 
single--photon detections \cite{fiura03}.
Moreover, knowledge of these few quantities is also sufficient
to determine the multimode, multipartite entanglement of
the state $\sig$. In fact, the fully symmetric $N$-mode block
$\sig_{\beta^N}$ can be again regarded as a state describing a mode
with CM $\gr\beta$ coupled with a fully symmetric $(N-1)$-mode block
$\sig_{\beta^{N-1}}$, and thus the $1\times(N-1)$ entanglement
within $\sig_{\beta^N}$ can again be computed by constructing the
corresponding equivalent two-mode state and evaluating its
entanglement. This scaling procedure can be iterated to determine all the
multimode entanglements existing between each mode and each fully
symmetric $K$-mode sub-block $\sig_{\beta^K}$ 
($K = 1, \ldots, N$). The
$1\times K$ entanglement between the single mode $\gr\alpha$ and any
fully symmetric $K$-mode partition $\sig_{\beta^K}$ of
$\sig_{\beta^N}$ can be determined in a similar way.
A crucial feature of this scaling structure of the multipartite
entanglement is that, at every step of the cascade, the $1 \times K$
entanglement is always equivalent to a $1 \times 1$ entanglement, so
that the quantum correlations between the different partitions of
$\sig$ can be directly compared to each other: it is thus possible
to establish a multimode entanglement hierarchy without any problem
of ordering. 

\begin{figure}[bt!]
 \includegraphics[width=6cm]{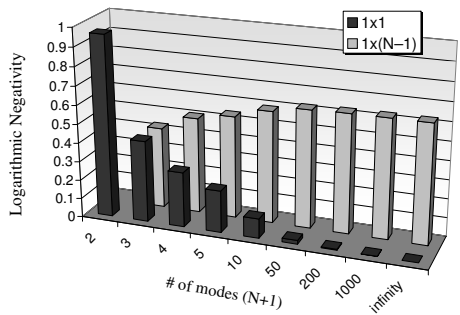}
 \caption{Scaling as a function of $N$ of the $1 \times 1$ and of the $1 \times (N-1)$ entanglement
 for a $(N+1)$--mode GHZ--type CV pure state ($b=1.5$).}
 \label{figscal}
\end{figure}

To illustrate the scaling
structure of multipartite entanglement in CV systems let us consider
a pure, $(N+1)$--mode fully symmetric Gaussian state of the form of \eq{fscm}.
Imposing the constraint of pure state ($\mu=1 \,
\Leftrightarrow \nu_- = \nu_{+^{(N+1)}}=1$), one obtains
$e_{i} = [1+b^2(N-1)-N -(-1)^{i}\sqrt{(b^{2} -1)(b^2(N+1)^2-(N-1)^2)}]/2bN$.
Such a state belongs to the class of CV GHZ--type states discussed in Ref. 
\cite{vloock03}. These multipartite entangled states are the
outputs of a sequence of $N$ beam splitters with $N+1$ single--mode
squeezed inputs \cite{vloock00,vloock03}. In the limit of infinite squeezing, these
states reduce to the simultaneous eigenstates of the relative positions and
the total momentum, which define the proper GHZ states of CV systems \cite{vloock03}.
The CM $\sig^{\rm GHZ}_{\beta^{N+1}}$ of this class of pure
states, for a given number of modes, depends only on the parameter
$b\equiv 1/\mu_\beta \ge 1$, which is an increasing function of the 
single-mode squeezing. Correlations between the modes are induced according 
to the above expression for the covariances $e_{i}$.
Exploiting our previous analysis, we can compute the entanglement between a single mode with reduced 
CM $\gr\beta$ and any $K$-mode partition of the remaining modes ($1 \le K \le N$), by determining
the equivalent two--mode CM $\sig^{eq}_{\beta\vert\beta^K}$.
The $1\times K$ entanglement quantified by the logarithmic negativity
$E_\N^{\beta\vert\beta^K}$ is determined by the smallest symplectic
eigenvalue $\tilde{n}_-^{(K,N)}$ of the partially transposed CM
$\tilde{\sig}^{eq}_{\beta\vert\beta^K}$.
\begin{figure}[t!]
\includegraphics[width=7.5cm]{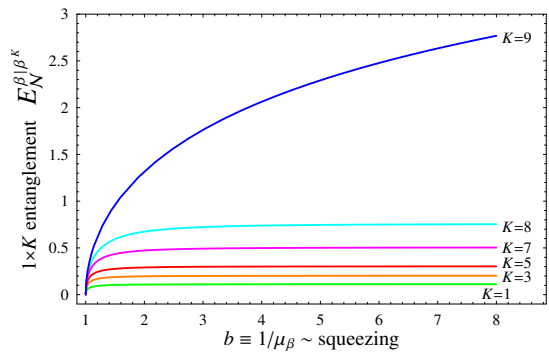}
\caption{(color online). Entanglement hierarchy for $(N+1)$-mode GHZ-type states ($N=9$).}
\label{fighz}
\end{figure}
For any nonzero
squeezing ({\em i.e.} $b>1$) one has that $\tilde{n}_-^{(K,N)}<1$, meaning
that the state exhibits genuine multipartite entanglement: each mode
is entangled with any other $K$--mode block, as first remarked in Ref. \cite{vloock03}. 
Further, the genuine multipartite nature of the entanglement can be precisely quantified 
by observing that $E_\N^{\beta\vert\beta^K} \ge
E_\N^{\beta\vert\beta^{K-1}}$, as shown in Fig.~\ref{fighz}. The
$1\times 1$ entanglement between two modes is weaker
than the $1\times 2$ one between a mode and other two modes, which is
in turn weaker than the $1\times K$ one, and so on with increasing
$K$ in this typical cascade structure. 
From an operational point of view, the signature of genuine multipartite entanglement
is revealed by the fact that performing {\em e.g.} a local measurement on a single mode
will affect {\em all} the other $N$ modes. This means that the quantum correlations
contained in the state with CM $\sig^{\rm GHZ}_{\beta^{N+1}}$ can be fully recovered
only when considering the $1 \times N$ partition.
In particular, the pure-state $1 \times N$ logarithmic negativity is, as
expected, independent of $N$, being a simple monotonic function of the
entropy of entanglement $E_V$ (defined as the von
Neumann entropy of the reduced single-mode state with CM $\gr\beta$).
It is worth noting that, in the limit of infinite squeezing ($b
\rightarrow \infty$), only the $1\times N$ entanglement diverges
while all the other $1\times K$ quantum correlations remain finite
(see Fig.~\ref{fighz}). Namely, $E_\N^{\beta\vert\beta^K}\!\!\big(\!\sig^{\rm
GHZ}_{\beta^{N+1}}\!\big)\overset{b\rightarrow\infty}{\longrightarrow}
-(1/2)\log\left[1- 4K/(N(K+1)-K(K-3))\right]$, which cannot exceed 
$\log\sqrt5 \simeq 0.8$ for any $N$ and for any $K<N$. 
At fixed squeezing, the
scaling with $N$ of the $1 \times (N-1)$ entanglement compared to the
$1\times 1$ entanglement is shown in Fig.~\ref{figscal} (we recall
that the $1\times N$ entanglement is independent on $N$). Notice
how, with increasing number of modes, the multipartite
entanglement increases to the detriment of the two-mode one 
which becomes distributed between all the modes.
We remark that this scaling occurs in any Gaussian states, either fully or  
partially symmetric, pure or mixed. For instance, this is the case
for a single--mode squeezed state coupled with a $N$--mode symmetric 
thermal squeezed state. 
The simplest example of a mixed state with genuine multipartite entanglement
is obtained from $\sig^{\rm GHZ}_{\beta^{N+1}}$ by tracing out some of the modes. 
Fig.~\ref{figscal}
can then also be seen as a demonstration of the scaling in such a
$N$-mode mixed state, where the $1 \times (N-1)$ entanglement is the
strongest one. Thus, with increasing $N$, the global mixedness can 
limit but not destroy the genuine multipartite entanglement between 
all the modes. This entanglement is
experimentally accessible by all-optical means
\cite{vloock03} and it also allows for a reliable ({\em
i.e.}~with fidelity ${\cal F}>1/2$) quantum teleportation between
any two parties \cite{vloock00}. Therefore, the quantification of
multipartite entanglement by measurements of purity, which,
as we have already remarked, can be experimentally implemented even 
without homodyning, leads to an accurate estimate of the multi-party 
teleportation efficiency and to direct control on the transfer of 
quantum information.

In conclusion, we have shown that multipartite quantum correlations
of Gaussian states of $1 \times N$ partitions
under symmetry are endowed with a scaling structure that 
reduces the problem to the analysis of the
entanglement of equivalent two--mode Gaussian states. Thanks to this
reduction, it is possible to determine exactly the logarithmic negativity
of the multimode states and to allow for a reliable experimental estimate
of the multipartite entanglement by direct measurements of global and local 
purities, without the need for the full recontruction of the covariance
matrix. Our results apply to many cases of practical interest. 
For instance, the entire class of bi-symmetric -- {\em i.e.} invariant under the
exchange of two given modes -- three-mode Gaussian states 
\cite{ngau2} has its multipartite entanglement completely quantified 
by the present analysis.
The generalization of the present approach for the quantification of 
multipartite CV entanglement to states with weaker symmetry constraints
and to $M \times N$--mode partitions (with $M>1$) awaits further study.

Financial support from INFM, INFN, and MIUR is acknowledged.

\noindent{\em Appendix: Proof of the symplectic
degeneracy.}\label{multi} We prove here the multiplicity of the
symplectic eigenvalue $\nu_-$ for the CMs $\sig_{\beta^N}$ and
$\sig$, asserted in Eqs.~(\ref{fssymp}) and (\ref{sympn}). We first
recall that, if $\Sigma=\{\nu_1,\ldots,\nu_N\}$ is the symplectic
spectrum of the CM $\sig$, then the $2N$ eigenvalues of the matrix
$i\Omega\sig$ are given by the set $\{\mp\nu_i\}$. Let us focus next
on the CM $\sig_{\beta^2}$: in the linear space on which the matrix
$i\Omega\sig_{\beta^2}$ acts, the eigenvector $v_-$ corresponding to
the eigenvalue $\nu_-$ reads
$v_-=(-i\frac{b-e_1}{\nu_-},-1,i\frac{b-e_1}{\nu_-},1)^{\sf T}$. Due to
the symmetry of $\sig_{\beta^{N}}$, any $2N$-dimensional vector $v$
of the form $ v\!=\!(0,\ldots0,\underbrace{-i\frac{b-e_1}{\nu_-},-1}_{{\rm
mode}\;i}, 0\ldots0,\underbrace{i\frac{b-e_1}{\nu_-},1}_{{\rm
mode}\;j},0,\ldots0)^{\sf T}$ ({\em
i.e.~}any vector obtained by taking $v_-$ in a couple of modes $ij$
and appending to it $0$ elements for all the other modes) is an
eigenvector of $i\Omega\sig_{\beta^N}$ with eigenvalue $\nu_-$. It
is immediate to see that one can construct $N-1$ linear independent
vectors of the above form, proving \eq{fssymp}. Clearly, an
analogous reasoning holds for the matrix $\sig$, proving \eq{sympn}.

\end{document}